\documentstyle[prb,multicol,aps,psfig,epsf]{revtex}
\begin{document}
\setlength{\parindent}{0em}
\draft
\title{Effect of spin glass order on magnetic polarons in
semimagnetic semiconductors}  
\author{A.L. Chudnovskiy, R. Oppermann, B. Rosenow}  
\address{Inst. f. Theoret. Physik, Universit\"at W\"urzburg, 
D--97074 W\"urzburg, F.R.Germany}  
\author{D.R. Yakovlev $^\ast$, U. Zehnder, W. Ossau}
\address{Physikalisches Institut, Universit\"at W\"urzburg, D--97074
W\"urzburg, F.R.Germany}
\date{\today}
\maketitle

\begin{abstract} 
A theory accounting for the specific features of magnetic polarons (MP) in the  
presence of spin glass order is presented. We derive and solve selfconsistent 
equations for i) the polaron magnetisation, ii) the thermodynamically averaged
carrier--spin, and iii) for the spin glass order parameter. The temperature 
dependence of these quantities is analysed in detail. The modification of the 
spin glass phase due to the presence of the exchange field of the carrier 
inside the magnetic polaron volume is investigated. The onset of spin glass 
order leads to a plateau--like flattening in the temperature dependence of the  
MP energy at low temperatures. It is found that solutions of spin glass 
equations are needed to optimally fit the experimental data of the temperature
dependence of the exciton magnetic polaron (EMP) energy in (Cd,Mn)Te. 
Moreover, the dynamical aspects of the MP formation are discussed. Our model 
predicts qualitatively different temperature dependences of the MP formation 
time in different dynamical scenarios.
\end{abstract}

\section{Introduction}

Diluted magnetic semiconductors (DMS) exhibit a variety of magnetooptical 
effects (giant Zeeman splitting, giant Faraday rotation etc.) which allow for  
the investigation of these materials with both standard magnetometric and 
optical techniques (for detailed review see Ref.\cite{Furdyna}). The DMS 
(Cd,Mn)Te has been studied intensively during the last decades \cite{land}$^-$
\cite{gaj}, its magnetic properties are determined by the spin $5/2$ of the 
Mn$^{2+}$--ions substituting Cd--ions randomly on the geometrically frustrated  
fcc lattice. Excitons are prominent quasiparticles in this material and probe 
locally the magnetic state of the system. Under suitable conditions they align 
the surroundig manganese spins and turn into polarons. The properties of these 
new excitations are determined by a variety of sample parameters like quality 
and structure of barrier surfaces. However, in all probes with sufficiently 
high manganese content (or at temperatures below 1 K) the disorder induced 
spin glass (SG) phase must be taken into account. We focus on this aspect of 
polaronic DMS physics.

The magnetically active manganese ions are coupled via an antiferromagnetic
interaction whose strength falls off rapidly with distance (the next--nearest 
neighbor (NNN) coupling is one order smaller than the nearest--neighbor (NN) 
interaction, which equals $-5.7 K$ \cite{rigaux}). By increasing the manganese 
concentration, clusters of magnetically coupled Mn--ions are formed. At 
concentrations above 20\% a percolating Mn--cluster emerges, exhibiting a 
transition into a spin glass phase at low (helium) temperatures, which is  
caused by the combination of geometric frustration and site randomness  
\cite{{villain},{rigaux},{oseroff}}. The SG phase is characterised by the 
order parameter $q=lim_{t\rightarrow\infty}\langle S(0)S(t)\rangle$, where $t$  
is the observation time, $S(0)$ and $S(t)$ denote spins at identical sites and  
at times $0$ and $t$, and brackets represent the thermodynamic average. 
In zero field $q$ is zero in the paramagnetic phase and assumes a nonzero 
value in the spin glass phase \cite{Fisher}.\\
Experimentally the SG phase transition has been observed by a cusp in the 
temperature dependence of the magnetic susceptibility and by the onset of 
irreversibility in the field cooled and zero field cooled measurements of the 
remanent magnetisation (for reviews see Refs.\cite{{wolff},{jang},{Fisher}}). 
The explicit use of spin glass equations for fitting experimental data of the 
magnetic polaron energy is however new and a key point of the present work. 

The manganese ions interact strongly with electronic states in the valence and 
conduction bands. In magnetic fields this leads to the giant Zeeman splitting 
of the band states, which extends up to tenths of meV. In the absence of an 
external magnetic field the formation of the magnetic polaron (MP) state -- 
a carrier with a cloud of magnetically polarised Mn spins which has a lowered  
energy -- becomes  possible. The dominant contribution to  this energy gain 
comes from the exchange interaction of manganese spins with the hole--spin. 
Magnetic polarons were investigated in detail for the case of carriers bound 
to a shallow impurity (for review see Ref.\cite{wolff}) and for excitons 
localised by compositional fluctuations \cite{8}.

Since the properties of the magnetic polaron are strongly influenced by the 
magnetic state of the system, one can expect that the formation of the spin 
glass phase should affect the properties of this quasiparticle. The onset of 
SG order should show up for example in the temperature dependence of the MP  
energy. The aim of the present work is to analyse theoretically the effects 
related to the interaction between the magnetic polaron and SG order, proving 
that SG effects are specific, hence distinguishable from any kind of 
antiferromagnetic clustering effects, and indeed required to explain 
qualitatively and quantitatively the new experimental data. 

The paper is organized as follows. In section \ref{sectheo} a model which 
accounts for the interaction between spin glass and magnetic polaron is 
presented and selfconsistent equations are derived for this model. An 
analytical analysis of these equations is presented in section \ref{secanalyt} 
followed by detailed numerical evaluations and a comparison of the theoretical 
results with the experimental data in section \ref{secnum}. Finally in section 
\ref{secdyn} an extended model, which takes into account dynamical aspects of 
magnetic polaron formation in the presence of spin glass order is presented.

\section{The Model and its Coupled Polaron Spin Glass Equations}
\label{sectheo}

In this paper we consider the magnetic part of the MP energy, which is 
directly influenced by the SG ordering of the Mn spin system 
\cite{{kavkav},{20}}. The exchange energy can also lead to the shrinking of 
the wave function of the polaron, thus contributing indirectly to the change 
of the kinetic energy of the MP. This contribution will be neglected 
throughout this paper. Furthermore, the contribution of the hole dominates 
the exchange energy of MP in (Cd,Mn)Te \cite{{8},{20}}. For simplicity we 
account for the hole contribution only, neglecting the small part of the 
exchange energy gained via coupling between electron and spin glass. These 
assumptions are surely justified for localised excitons.

We adopt here the Villain model of Ising pseudospins \cite{{villain},{rigaux}}.  
In Ref.\cite{rigaux} detailed arguments were given in favor of this model and 
its applicability to (Cd,Mn)Te. The lower critical dimension for its Ising SG 
transition being smaller than three together with the fact that the  
Dzyaloshinskii--Moriya interaction, which would also turn the Heisenberg into 
an Ising SG (provided this coupling is isotropic itself; anisotropy would lead 
to special universality classes \cite{stein}), is too small, supports the 
above model description. In the model considered here each pseudospin belongs to
a tetrahedron with Mn spins on the corners. Due to the nearest neighbor 
antiferromagnetic interaction between the Mn spins the tetrahedron is 
frustrated and its ground states can be characterised by a mixed product 
$\xi_{ijk}={\bf\sigma_i\cdot(\sigma_j\times\sigma_k)}$ of three neighboring 
spins which can assume only two opposite values. The normalised value of this 
product $S=\xi/|\xi|$ can be $1$ or $-1$ and is hence considered as an  
Ising pseudospin.

The interaction between hole-- and Mn--spins will be described by the 
following model. Consider a percolating cluster with $N$ pseudospins in
\nolinebreak[4] (Cd,Mn)Te with random frustrated interaction between the spins 
leading to spin glass order at low temperatures. We assume the interaction 
between Ising pseudospins to be of infinite range. The hole spin of the 
exciton interacts with a part of $N_1=\alpha N$ pseudospins of the entire spin 
cluster. In this way the finite range of this exchange coupling is taken into 
account. We represent this Kondo--like coupling by $-(\kappa/N_1) 
s^z\sum_{i=1}^{N_1}S_i$. Here $\kappa$, $s^z$, and $S_i$ denote the effective  
exchange constant, $z$-component of the hole spin, and the Ising pseudospin at 
site $i$ respectively. The "exchange box"--model is assumed, i.e. the wave
function of the hole is constant inside the magnetic polaron and zero outside 
\cite{potbox}. Our assumption implies $|\Psi(r)|^2=1/N_1$ inside the polaron 
and zero outside. Therefore the Hamiltonian of the model is given by
\begin{equation}     
H=-\frac{\kappa}{N_1}
s^z\sum^{N_1}_{i=1}S_i -\sum_{(i,j)}^N J_{ij}S_iS_j. 
\label{Hs}
\end{equation} 
The term $-\sum_{(i,j)}^NJ_{ij}S_iS_j$ describes the random interaction 
between spins of the spin glass with gaussian--distributed random $J_{ij}$ of 
width $J$ and the convention that $(i,j)$ counts pairs $i,j$ only once and 
only with $i\neq j$.

To investigate the physical effects which result from this model we express 
the partition function employing the replica--trick, perform the configuration 
average over the frustrated spin couplings, and after the decoupling procedure    
by use of the Hubbard--Stratonovich transformation obtain the saddle point
equations for the model (for the detailed description of this technique see 
Ref.\cite{jang}). The partition function of the model is thus obtained as
\begin{equation}
\langle    Z^n\rangle =\prod\int
P(J_{ij})dJ_{ij}\sum_{S_i,s=\pm 1}
\exp\left[\sum_{a=1}^n\beta\left\{\frac{\kappa}{N_1}
s^{a}\sum_{i=1}^{N_1}S_i^{a}+\sum_{i,j}^N J_{ij}  
S_i^{a}S_j^{a}\right\}\right], 
\label{statsum}
\end{equation} 
with $\beta\equiv 1/T$. $P(J_{ij})$ denotes the Gaussian distribution of 
couplings, the superscripts $a$ relate to replicas. Here and in what follows 
the superscript $z$ will be dropped for the hole--spin $s$.  

The Kondo-term $-(\kappa/N_1)s\sum_{i=1}^{N_1}S_i$ is decoupled with the help 
of the identity 
\begin{equation}
s\sum_{i=1}^{N_1}S_i=\frac{1}{2}((s+\sum_{i=1}^{N_1}S_i)^2-s^2- 
(\sum_{i=1}^{N_1}S_i)^2).
\end{equation}
The spin glass term in the exponent of the partition function is identical to 
the hamiltonian of the wellknown Sherrington-Kirkpatrick (SK) model (see 
\cite{{jang},{sk}}). Thus the decoupling of this term is completely equivalent 
to that of the SK model.

The whole decoupling procedure transforms the partition function into  
\begin{eqnarray} 
\nonumber 
\langle Z^n\rangle &= &\sum_{s,S_i}\int\prod dV_a dM_a dR_a dQ^{ab}                
\exp\left[\beta \sum_{a=0}^n\left\{-\frac{\kappa}{2}\left(N_1 
R_a^2+N_1M_a^2+1/N_1V_a^2\right)\right.\right.\\  
\nonumber  
& & \left.\left. +\kappa (R_a+iV_a/N_1)s^{a}+\kappa (R_a+iM_a)
\sum_{i=1}^{N_1}S_i^{a}\right\}\right. \\
& &\left.+\sum_{a,b}^n\left\{-\beta^2\frac{N J^2}{4}(Q^{ab})^2 
+\beta^2\frac{J^2}{2}\sum_{i=1}^NQ^{ab}S_i^{a}S_i^{b}\right\} 
\right], 
\end{eqnarray}\label{Zdec} 
where the fields $R_a, V_a, M_a$ facilitate the decoupling of the Kondo--like 
interaction, while the field $Q^{ab}$ does the job for the spin glass 
interaction. 

Now we define the replica--symmetric and spatially homogeneous saddle point    
Ansatz for the decoupling fields by
\begin{eqnarray}
R_{a}&=&\eta \\
M_{a}&=&im\\
V_{a}&=&ip \\
Q^{ab}&=&q, \mbox{for $a\neq b$}.
\end{eqnarray}\label{sprmp}

For the fields $Q^{ab}$, replica--diagonal and replica--offdiagonal components 
are distinguished in general. The replica--offdiagonal components are related  
to the SG order parameter, whereas the replica diagonal components are 
noncritical for the Ising spin glass model \cite{jang}. After the substitution 
of the saddle--point Ansatz in the partition function the free energy of the 
system is obtained as follows

\begin{eqnarray}
\nonumber
\beta F&=&\frac{1}{2}\frac{N_1}{N}\beta\kappa(\eta^2-m^2-(p/N_1)^2) 
+\frac{\beta^2J^2}{2}\left[(q-1)^2 \right. \\
&& -\frac{N_1}{N}\int^G_z\ln\cos h\left[\beta(J\sqrt{q}z 
+\kappa(\eta-m))\right] -\frac{N-N_1}{N}\int^G_z 
\ln\cosh\left[\beta J\sqrt{q}z\right],
\end{eqnarray}\label{freeen}
where
\begin{equation}
\int^G_z f(z)\equiv \frac{1}{\sqrt{2\pi}}\int dz e^{-z^2/2} f(z).
\end{equation}

The free energy becomes extremised by $\eta$, $p$, $m$ and $q$ satisfying the 
following saddle point equations

\begin{equation} 
\eta=p/N_1+m
\label{R}\end{equation}
\begin{equation}
p=\tanh\left[\beta\kappa m\right]
\label{p}\end{equation}
\begin{equation}
m=\int^G_z \tanh\left[\beta(J\sqrt{q}z+\kappa        
\frac{p}{N_1})\right] 
\label{m}\end{equation} 
\begin{eqnarray} 
\nonumber q=\frac{N_1}{N}\int^G_z \tanh^2\left[\beta (J\sqrt{q}z  
+\kappa \frac{p}{N_1})\right]&&\\
+\frac{N-N_1}{N}\int^G_z \tanh^2\left[\beta J
\sqrt{q}z\right]
\label{q}\end{eqnarray}

The value $q=\frac{N_1}{N}\langle S^aS^b\rangle_{ins}+\frac{N-N_1}{N}\langle 
S^aS^b\rangle_{outs}$ ($a\neq b$) is the weighted average of Edwards--Anderson 
order parameters inside and outside the polaron respectively \cite{jang}. 
$p=\langle s^{a}\rangle$ represents the thermodynamically averaged spin of the 
hole and $m=1/N_1\langle\sum_{i=1}^{N_1}S_i^a \rangle$ denotes the average 
magnetisation in the polaron. The equations (\ref{p})--(\ref{q}) describe the 
whole phase diagram of the model in the static mean field approximation. From  
the qualitative analysis of these equations three characteristic temperature  
ranges can be identified.\\
(i) At high temperatures $T>\kappa/\sqrt {N_1}$ the equations have only 
trivial solutions. The system is paramagnetic and thermal fluctuations 
suppress the magnetic polaron formation. \\ 
(ii) At $\kappa/\sqrt{N_1}>T>J$ the magnetic polaron is formed but the Mn 
system remains in the paramagnetic phase.\\  
(iii) At $T<J$ the SG order is expected to show up in the Mn system. Hence the 
value $\kappa/\sqrt{N_1}$ determines the temperature of the MP formation. In 
the experiments the MP is formed in the paramagnetic temperature regime, which 
implies the inequality $\kappa/\sqrt{N_1}>J$.

\section{Temperature dependences of magnetisation in the polaron and of the 
spin glass order parameter in the vicinity of the phase transition} 
\label{secanalyt}

Here we examine the dependence of the SG order parameter $q$ and of the 
cluster magnetisation $m$ on $T-J$, where $J$ corresponds to the transition 
temperature of an independent SG model. The main difference between the  
present case described by equation (\ref{q}) and the analogous equation for  
the independent spin glass \cite{jang} is caused by the presence of the 
exchange field 
\begin{equation} h_{exch}=\kappa\frac{p}{N_1} 
\end{equation}  
which the localised hole exerts on the pseudospin. It is worth noting that the  
influence of the exchange field on the magnetisation in the polaron and on the 
SG order parameter $q$ is analogous to the influence of an external magnetic 
field localized in the volume of the magnetic polaron (see Eqs.(\ref{m}) and 
(\ref{q})). The nonzero magnetic field competes with the exchange interaction 
between Mn spins and destroys partially their ordering. This results in the 
smearing of the phase transition and lowering of the (smeared) transition 
temperature \cite{jang}. The transition should still show up in a change of 
the temperature behavior of the magnetization $m$ and of the SG order parameter 
$q$. In order to see these expected changes we analyse the temperature 
behavior of the quantities $m$ and $q$ in the vicinity of the point $T=J$. The 
case when the polaron constitutes a fixed part of the Mn spin system is 
considered, i.e. $\alpha=\frac{N_1}{N}$ fixed, $0<\alpha <1$. We expand the 
right--hand sides of Eqs.(\ref{m}), (\ref{q}) for small $q$ up to quadratic 
terms. After integration over the random field $z$ the equation for $q$ becomes
\begin{equation}
2\beta^4J^4q^2+(1-\beta^2J^2(1-4C))q-C=0,
\end{equation}\label{critq}
where $C\equiv\alpha\tanh^2(\beta h_{exch})$. The physical solution for $q$ 
reads
\begin{equation}
q=\frac{\beta^2J^2-1-4\beta^2J^2C+
\sqrt{(1-\beta^2J^2)^2+8\beta^2J^2C}}{4\beta^4J^4}.
\end{equation}
In the region $|1-\beta^2J^2|\gg\beta^2J^2C$ the behavior of $q$ changes (for 
$C=0$ nonanalytically) at $\beta J=1$. For $T>T_c$ (i.e. in the paramagnetic 
phase) 
\begin{equation} 
q\approx C\approx\frac{\alpha h_{exch}^2}{T_c^2}   
\left(1-2\theta\right),
\end{equation} 
where $\theta\equiv T/T_c-1$. In the SG-phase, $T<T_c$, 
\begin{equation}
q\approx 2\frac{h_{exch}^2}{T_c^2}\left(1+2\theta\right)-\theta.  
\end{equation}    
The dependence $q(\theta)$ is linear both above and below $T_c$, but one 
observes different slopes in the paramagnetic phase and in the SG phase. 
Similar behavior is observed in the magnetisation $m$. We find 
\begin{equation} 
m\approx\left\{
\begin{array}{cc} 
\frac{h_{exch}}{T_c}\left(1-\theta\right)&  \mbox{for $T>T_c$},\\ 
\frac{h_{exch}}{T_c}-\alpha\frac{h_{exch}^3}{T_c^3}\left(1+25\theta\right)& 
 \mbox{for  $T<T_c$}.  
\end{array}\right.  
\label{mt} 
\end{equation} 
The linear dependence of $m$ on $\theta$ becomes very weak below $T_c$, 
i.e. in the SG phase.

The solutions of the saddle point equations describe also the smearing of the 
otherwise sharp phase transition as caused by the finite exchange field in the 
magnetic polaron. The smearing is efficient around the point $T_c=J$ within  
the temperature interval given by $|1-\beta^2J^2|\ll\beta^2J^2C$. Beyond this 
interval the power law behavior of $m$ and $q$ characteristic of the 
transition in the pure spin glass is retrieved.
\newpage

\section{Numerical saddle--point solutions and comparison with experiments}
\label{secnum}       

In order to obtain a detailed description of the physical behavior we solved 
the saddle-point equations (\ref{p})--(\ref{q}) numerically. It is convenient 
to define a composite SG order parameter according to 
$q=\alpha q_1+(1-\alpha)q_2$, where
\begin{equation}
q_1\equiv\int_G dz\tanh^2\left(\beta(J\sqrt{q}z+\kappa\frac{p}{N_1})\right),
\end{equation}\label{q1}
\begin{equation}
q_2\equiv\int_G dz\tanh^2\left(\beta J\sqrt{q}z\right).
\label{q2}
\end{equation}
describe spin glass order inside ($q_1$) and outside ($q_2$) the magnetic 
polaron, respectively.\\ 
A qualitative understanding of the coupled selfconsistent solutions involving 
polarisation $p$, magnetisation $m$, and the SG order parameters $q_1$, $q_2$ 
can be obtained as follows. Fig.\ref{fig1} displays the magnetisation, which 
is determined from Eq.(\ref{m}), and the SG order parameters. The thermal 
behavior of $m$ and $q_1$ is strongly affected by the coupling $\kappa_1\equiv 
\kappa/N_1$ which controls the interaction energy of a pseudospin with the 
hole--spin. For a large value $\kappa_1=3.14 meV$ (curve 1 in Fig.\ref{fig1}) 
the magnetisation $m$ decreases monotonuously with temperature and the phase 
transition appears strongly smeared by a large exchange field (which keeps 
$q_1$ at high values due to non spin glass contributions and 
thus inhibits its critical decay at the zero field freezing 
temperature). At smaller $\kappa_1=0.628 meV$ (curve 2 in Fig.\ref{fig1}) 
instead, $m(T)$ displays a maximum near the almost sharp SG transition  
signalled by the rather steep decay of both SG order parameters $q_1$, 
within the MP, and $q_2$. This is in accordance with the analytical 
solution given in Eq.(\ref{mt}).
For fixed temperature, $m$ and $q_1$ grow monotonuously with $\kappa_1$.   
The temperature dependence of the average hole--spin $p$ at two different 
values of the interaction energy $\kappa$ between hole--spin and pseudospins 
within the MP is shown in Fig.\ref{figp-t}. The change of $\kappa$ affects the 
thermal behavior of the polarisation $p(T)$ in a similarly strong way as a 
$\kappa_1$--variation alters the magnetisation $m(T)$. In turn, the spin glass 
order parameter $q_2$ outside the magnetic polaron depends very weakly both on 
$\kappa$ and $\kappa_1$. Fig.\ref{fig1} shows that with increasing $\kappa_1$ 
the temperature dependence of $q_1$, which resembles the $q_2$--curve for 
small $\kappa_1$ (almost sharp SG transition), changes over and approaches the 
magnetisation curve $m(T)$ in the large $\kappa_1$--regime. 

The total number $N$ of pseudospins enters in all formulae only in the form
$N_1=\alpha N$. Since, additionally, the distinction between magnetic order 
inside the MP and the one within the remaining spin glass domain (not 
exposed to the exchange field) will influence the quantitative agreement 
between theory and experiment, it is convenient to analyse thermal properties
as a function of $N_1$ and $\alpha$. $N_1$ together with $\kappa$ determines 
the exchange energy $\kappa_1$ on a pseudospin. Changing $N_1$, at fixed 
$\kappa$ and $\alpha$, is thus equivalent to tuning $\kappa_1$, which was 
defined in the preceding paragraph. The parameter $\alpha$ determines the 
relative extension of the polaron and enters in the SG order parameter via the 
relation $q=\alpha q_1+(1-\alpha)q_2$. Large values of $\alpha$ correspond to 
a strong influence of the polaronic exchange field on the spin glass order 
because $\alpha$ is the weight of the $q_1$--contribution to the composite SG 
order parameter $q$. As $\alpha$ decreases, the peak $m(T)$ shifts to higher 
temperatures and becomes more pronounced (see Fig.\ref{figalpha}).

Measurements of the magnetic-field-induced circular polarisation degree of 
exciton luminescence (reflecting the magnetic susceptibility \cite{6}) and of 
the magnetic polaron energy (determined by selective excitation of localised 
excitons \cite{8}) against temperature are displayed in Fig.\ref{figmpe}.    
The polarisation degree shows a cusp at $T=8K$ characteristic of the spin 
glass phase transition \cite{{kudinov},{7}}. The cusp temperature coincides 
with the SG transition temperature, which was obtained for this manganese 
concentration by other measurements (see Ref.\cite{{wolff},{oseroff}}). 
The experiments on the temperature dependent EMP energy (solid circles in 
Fig.\ref{figmpe}) do not show any peculiarity at the spin glass transition 
temperature $T_c=8K$. However, a plateau is observed at lower temperatures. 
We argue that this plateau results from the onset of SG order inside the 
polaron. The plateau extends over a temperature range lying below the 
cusp temperature of the zero field spin glass, because of the large 
exchange field inside the MP. 

In order to determine the parameters of our model we fitted the experimentally 
known temperature dependence of the exciton magnetic polaron (EMP) energy in 
a Cd$_{0.67}$Mn$_{0.33}$Te epilayer.  
The exchange part of this energy is calculated according to  
\begin{equation} 
\epsilon=\kappa mp \\ 
\label{energy1} 
\end{equation} 
We fix the value of $J=8K$ (the cusp temperature in the circular polarisation  
data) whereas the effective exchange interaction $\kappa$ between a hole spin 
and pseudospins (tetrahedrae of Mn spins) in the polaron is chosen as a fit 
parameter. We also vary the parameters $N$ and $\alpha$ to obtain maximal 
agreement of the calculation with the experimental points. 
The best fit, given by the solid line in Fig.\ref{figmpe}, is obtained with 
$\alpha=1$, $N=20$, $\kappa=40.5 meV$, which is already in a rather good 
agreement with the experimental data. This justifies our assumption that the 
magnetic contribution to the EMP energy dominates thermal variations and is
hence responsible for the main features of its temperature dependence. The 
numerical calculations indicate that for these parameters the hole spin is 
completely polarised, which means $p=1$ for all temperatures under 
consideration. Thus the temperature dependence of the EMP--energy reflects the 
polarisation degree of the Mn spins in the region of the excitonic wave 
function. Calculating the exchange field ($\sim\kappa/(\alpha N)$) in the 
polaron we obtain a value of about 4.7T, which is in reasonable agreement with 
the value 5.9T obtained by other experimental techniques \cite{6}. 
If we take into account that $N$ is the number of {\em pseudo}spins 
representing tetrahedrae of Mn spins, we can as well estimate the number of 
Mn spins inside the polaron to be about 60, which is roughly a factor of 2 
smaller than the value estimated from experiment \cite{{wolff},{6}}.

Despite the rather good agreement between experiment and the model description 
discussed so far, there are two reasons why the present situation is not yet
completely satisfactory. On one hand, $\alpha=1$ corresponds to a polaron
extending through the whole spin glass, on the other, it is known that, 
besides the percolating spin glass cluster, there exist small clusters of 
antiferromagnetically coupled manganese spins. It turned out that these two 
problems are connected, since an extended model containing bulk spin glass 
order {\it and} clusters of a small number of antiferro--coupled Mn spins 
allows to describe the experimental data with a more realistic value of 
$\alpha$. 

We model the temperature dependent polarisation of the small Mn--spin clusters 
by use of the wellknown modified Brillouin function \cite{gaj} commonly used 
to fit the experiments on Zeeman splitting of excitons in (Cd,Mn)Te 
\cite{wolff}. This yields
\begin{equation}
m_{cl}=xS_{eff}B_{5/2}\left(\frac{\frac{5}{2}\mu B_{exch}}{T+T_{eff}}  
\right).
\label{mcl}\end{equation}
Here the effective spin $S_{eff}<5/2$ and the effective temperature $T_{eff}$ 
reflect the antiferromagnetic coupling of Mn--spins (see Ref.\cite{gaj}); 
$x=0.33$ stands for the manganese concentration, and $\mu$ denotes the Bohr 
magneton. In the argument of the Brillouin function we insert instead of an  
external magnetic field the exchange field $B_{exch}$ between hole spin and 
Mn--spins in the MP. For the manganese concentration of 33\% the exchange 
field is $B_{exch} \approx 5.9 T$ \quad\cite{6}.

The magnetic part of the EMP energy is now given as the sum of the 
contribution originating in spin glass order and a second term representing 
the effect of small clusters; the total magnetic EMP energy then becomes
\begin{equation}
E_{EMP}=\frac{N_1}{N_1+N_{cl}}\kappa pm   
+ \frac{N_{cl}}{N_1+N_{cl}}\kappa_0 pm_{cl}, 
\label{energy2}\end{equation}
where $N_{cl}$ denotes the number of spins within small clusters in the MP. 
The exchange constant $\kappa_0=880 meV$ is known from experiment \cite{gaj}.

We fitted the experimental data with the form (\ref{energy2}). Now, in 
addition to $\kappa$, $\alpha$ and $N_1$, we face three fit parameters more. 
These are the number of spins in small clusters $N_{cl}$ and the parameters of 
the modified Brillouin function $S_{eff}$ and $T_{eff}$.
The aim of our procedure is now to prove that \\
i) it is possible to fit experimental data with allowance for the additional
presence of small clusters, keeping all parameter values within the 
physically meaningful regime, together with\\ 
ii) the plateau in the temperature dependence as a genuine effect of the 
spin glass. This plateau cannot be obtained selfconsistently from the model of 
clusters of antiferromagnetically coupled spins only. The fitting procedure 
is illustrated by Fig.\ref{figfit}. 

As mentioned above the searched for function $E_{EMP}(T)$ is determined by the 
temperature dependences of the magnetisations $m$ and $m_{cl}$ because the 
hole--spin is completely polarised in the entire temperature region under 
consideration. 

For bulk--like real systems one expects a nonzero portion of the spin glass
cluster to be out of reach of the exciton wavefunction, hence outside the
polaron. Therefore the real physical situation corresponds to $\alpha<1$. 
The ratio $N_1/N_{cl}$ defines the relative contribution of the SG and of 
small clusters to the EMP energy. The lack of experimental information on 
the precise value of $\alpha$ and $N_1/N_{cl}$ forced us to analyse fits with 
different values of $\alpha$ for the interval from 0.3 up to 0.7 and for 
$N_1/N_{cl}$ of order 1. The results for the fits were remarkably insensitive 
to the variation of $\alpha$ at least within this regime. In the fit presented 
in Figs.\ref{figmpe} (dash-dotted line) and \ref{figfit} (solid line) we fix 
$\alpha=0.5$ and $N_1=N_{cl}$, which we consider to be a good choice for the 
real system under consideration. While these relations were not rigorously
dictated from experiment, the insensitivity of the fits with respect to
variations of these two parameters over a wide range guarantees that the 
fit, given for the above choice, will describe the real system.

Having chosen $\alpha$ and $N_1/N_{cl}$, we are left effectively with four 
fit parameters, $\kappa$, $N_1$, $S_{eff}$, and $T_{eff}$, which remain to 
be determined. In the following we define the four relations employed here 
to determine these parameters unambiguously. 

The first three of these conditions are the principal experimental features,
which need to be reproduced by the fit curve: \\
i) the absolute plateau--height at low temperatures ($\approx 28 meV$),\\
ii) the high temperature end of the plateau ('plateau--ridge') at 
$\approx 3 K$, and \\
iii) the decay of the magnetic EMP energy at higher temperatures, 
as seen in the experiment way above the spin glass freezing.\\ 
The fourth condition is concluded from the following consideration:
The plateau ridge is strongly connected with the position of the peak 
caused by the SG contribution (dashed line in Fig.\ref{figfit}), 
which itself is determined by the value $\kappa_1=\kappa/N_1$. Keeping the 
position of the peak as close as possible to the plateau ridge, condition iv) 
results and we obtain a single value $\kappa_1$ for each $\kappa$; 
hence the parameter $N_1$ is now completely determined by the value of 
$\kappa$.  

The matching of the plateau's high temperature end with the experimental 
data is achieved by tuning the form of the contribution of small clusters 
(by means of the parameters $T_{eff}$ and $S_{eff}$) for each $\kappa$. 
This requirement provides a relation between $\kappa$, $S_{eff}$, and 
$T_{eff}$.

The absolute value of the plateau is defined by the parameters $S_{eff}$ and 
$\kappa$, whilst the role of $T_{eff}$ and $N_1$ is negligible. Thus the 
absolute plateau height of the energy curve yields a relation between 
$\kappa$ and $S_{eff}$. 

Finally the form of the decay at high temperatures as seen in the experiments
provides one more relation between the parameters $\kappa$, $S_{eff}$, 
$T_{eff}$, and $N_1$. 

Fig.\ref{figfit} shows the form of the SG contribution (dashed line) and that 
of the small cluster contribution (dashed-dotted line) to the EMP energy 
together with the resulting combined curve of the temperature dependence 
(solid line). Note that the form of the small cluster contribution is concave. 
It follows that the plateau results from the SG contribution to the 
EMP energy.

One can try to fit the same experimental data with the modified Brillouin 
function alone. However, in this case one should take the parameters obtained 
from the Zeeman splitting fit in Cd$_{0.67}$Mn$_{0.33}$Te, since now the 
Brillouin function describes the whole spin system and not only the part of 
small clusters as above. If one does so, the Brillouin fit does not coincide 
with the experimental data. Agreement can only be achieved by using Brillouin 
function fit--parameters very different from those obtained by the Zeeman 
splitting. For this reason such a procedure does not result in a consistent 
description of the complex magnetic behavior of (Cd,Mn)Te. 

In conclusion, the fit described above shows that the main qualitative 
feature, which is the plateau in the temperature dependence of the EMP energy, 
is the specific result of the spin glass order in the polaron.  
The plateau occurs at a temperature lower than the bulk SG transition 
temperature because of the high exchange field in the polaron.

\section{Dynamical aspects of magnetic polaron formation}\label{secdyn}

In this section we suggest a phenomenological model for the dynamics of the
MP formation. The approach presented here aims at taking into account the 
contribution of the SG-phase. Thus we do not consider the nonmagnetic 
contributions to the MP formation (prelocalisation), the shrinking of the 
hole--wavefunction during MP--formation and the microscopic mechanisms of the 
spin-- and energy relaxation (a description of these subjects is given in 
Ref.\cite{{dietl},{20},{21}}).

We suppose the changes of the averaged spin of the hole and the magnetisation 
in the polaron to be slow in time in comparison with the thermodynamic 
fluctuations so that, at each instant of time, we can define the state of the  
system by thermodynamically averaged quantities (magnetisation, average 
hole--spin etc.). Then the formation of the MP can be considered as an 
evolution of MP parameters $m$ and $p$ in time from zero values at $t=0$ which  
is the trivial solution of the static mean field equations (\ref{p} and 
\ref{m}) towards the nontrivial static solutions at $t\rightarrow\infty$. 
This evolution is governed by an additional dynamical part. Hence we propose 
the following phenomenological set of equations
\begin{equation}  
\ddot{p}+\gamma \dot{p} + \Omega^2 p=\Omega^2 \tanh\left[\beta\kappa   
m\right], 
\label{pdyn}  \end{equation}   
\begin{equation}    
\ddot{m}+\gamma \dot{m} + \Omega^2 m=\Omega^2\int_G
dz\tanh\left[\beta(J\sqrt{q}z+\kappa\frac{p}{N_1})\right].
\label{mdyn}\end{equation}

This form of the dynamical part contains only the leading time derivatives and 
is linear in $p$ and $m$. An alternative form containing only first time 
derivatives does not yield a time evolution under the starting conditions 
$p=0$, $m=0$. The Eqs.(\ref{pdyn}), (\ref{mdyn}) describe the dynamics of 
nonlinearly coupled harmonic oscillators with damping. The characteristic 
frequency $\Omega$ is the oscillation frequency, while $\gamma$ characterises 
the damping.

We assume $q$ to satisfy equation (\ref{q}) with time--dependent $p$ and $m$.  
This assumption is consistent with the definition of $q$ as Edwards-Anderson 
order parameter. The equilibration time for $q$ is of order $1/T$  --  the 
characteristic time of thermodynamic relaxation processes at this temperature. 
For the suggested model this time is shorter than the characteristic evolution 
time of $p$ and $m$.

The dynamics of the MP formation is described by the solution of equations  
(\ref{pdyn}), (\ref{mdyn}), (\ref{q}) with the initial conditions $p(t=0)=0, 
m(t=0)=0$. We can obtain an analytical solution for the initial stage of the 
MP formation where $\beta\kappa m\ll 1, \beta\kappa p/N_1 \ll 1$ so that we 
may expand in these values up to the linear terms. In the paramagnetic phase 
$q$ is nonzero only due to the exchange field of the MP. The relevant values 
of $z$ in the integral of Eq.(\ref{mdyn}) lie in the region 
$J\sqrt{q}z\ll\kappa p/N_1$. Hence we neglect $q$ in Eq.(\ref{mdyn}) and 
obtain for the initial step of MP-formation $t\ll min(1/\gamma, 1/\Omega)$ 
in the paramagnetic phase ($T>J$)
\begin{equation} 
\ddot{p}+\gamma  \dot{p} + \Omega^2 p=\Omega^2  \beta\kappa  m,  
\label{pdynpm}
\end{equation} 
\begin{equation}    
\ddot{m}+\gamma \dot{m}+\Omega^2 m =\Omega^2\beta\kappa\frac{p}{N_1 }.  
\label{mdynpm} 
\end{equation}
The nontrivial solution of this system of linear differential equations 
with the initial conditions $m(0)=0, p(0)=0$ is    
\begin{equation}
m=Ce^{-\gamma t/2} \sinh\left[\frac{t}{2}\sqrt{\gamma^2+        
4\Omega^2\left(\beta\kappa/\sqrt{N_1}-1\right)}   \right], 
\end{equation}\label{solmdynpm}                
\begin{equation}
p=C\sqrt{N_1}e^{-\gamma t/2}\sinh\left[\frac{t}{2}\sqrt{\gamma^2+ 
4\Omega^2\left(\beta\kappa/\sqrt{N_1}-1\right)} \right].
\label{solpdynpm}
\end{equation}
One obtains a nonoscillating solution, which corresponds to the formation of 
the polaron only under the condition $\beta\kappa/\sqrt{N_1}-1 >0$. The same 
inequality was obtained in the end of section \ref{sectheo} from the static  
selfconsistency equations as a necessary condition for the polaron formation.  
The constant $C$ and the characteristic frequencies $\gamma$ and $\Omega$ 
could be inferred from the comparison with experiment.

Let us now consider the case of spin glass order. In the vicinity of the phase 
transition the integral in Eq. (\ref{mdyn}) is dominated by $\beta J\sqrt{q}z \ll 1$.  
Since $\beta\kappa p/N_1 \ll 1$, we have 
\begin{eqnarray}                
\nonumber
\int_G dz\tanh\left[\beta(J\sqrt{q}z+\kappa \frac{p}{N_1}) 
\right]& \approx & \tanh(\beta\kappa p/N_1)\left(1-\int_G dz      
\tanh^2(\beta J\sqrt{q}z)\right) \\ 
& \equiv & \tanh(\beta\kappa p/N_1) (1-I(q)). 
\end{eqnarray} 
At times $t\ll 1/\gamma$ the condition $J\sqrt{q}z\gg \kappa p/N_1$ is 
fulfilled for the relevant domain of $z$ in the integral. Then we can set 
$I(q)\approx q$ with $q$ evaluated from Eq.(\ref{q}) with $p=0$. Expanding 
$tanh(x)\approx x$ the following equations are obtained 
\begin{equation} 
\ddot{p}+\gamma \dot{p} + \Omega^2 p =\Omega ^2 \beta\kappa m,  
\label{pdynsg}\end{equation} 
\begin{equation}    
\ddot{m}+\gamma \dot{m} + \Omega^2  
m=\Omega^2\beta\kappa\frac{p}{N_1}G(q), 
\label{mdynsg}\end{equation}
where $G(q)=1-I(q)\approx 1-q$. The solutions of these equations read
\begin{equation}  
m=Ce^{-\gamma t/2}\sinh\left[\frac{t}{2}                
\sqrt{\gamma^2+ 4\Omega^2\left(\beta\kappa\sqrt{G(q)/N_1}-1\right)} 
\right],    
\end{equation}\label{solmdynsg} 
\begin{equation}
p=C\sqrt{N_1/G(q)}e^{-\gamma t/2}\sinh\left[\frac{t}{2} 
\sqrt{\gamma^2+ 4\Omega^2\left(\beta\kappa\sqrt{G(q)/N_1}-1\right)} 
\right].
\label{solpdynsg}
\end{equation} 
>From $G=1-q<1$ it follows that the spin glass order slows down the magnetic 
polaron formation.

The time dependence of the exchange part of the MP energy at the initial stage 
of MP formation (Eq.(\ref{energy1})) is given by 
\begin{equation} 
\epsilon (t)\propto \sqrt{N_1/G(q)}e^{-\gamma t}         
\sinh^2\left[\frac{t}{2}\sqrt{\gamma^2+   
4\Omega^2\left(\beta\kappa\sqrt{G(q)/N_1}-1\right)} \right],
\label{eemp-t}
\end{equation}
where 
\begin{equation}
G(q)\equiv \left\{\begin{array}{ll} 1 & \mbox{in paramagnetic phase}\\
  1-I(q) & \mbox{in spin-glass phase}.
\end{array}
\right.
\end{equation}

Thus the MP formation depends exponentially on time. Representing the 
magnetic part of the MP energy as $\epsilon (t)\propto \exp(t/\tau_f)$, we 
obtain the following temperature dependence for the MP formation time
\begin{equation}
\tau_f\equiv \left(\sqrt{\gamma^2+
4\Omega^2\left(\beta\kappa\sqrt{G(q)/N_1}-1\right)} 
-\gamma\right)^{-1}
\label{tauf}\end{equation}

The introduced model allows one to discriminate two limiting cases of the MP 
formation dynamics: the case of "strong dissipation" $\gamma \gg \Omega$ and 
the case of "weak dissipation" $\Omega \gg \gamma$. The expressions for the 
MP formation time in these two limit cases read
\begin{equation}
\tau_f \approx \left\{ 
\begin{array}{lr}
\left[\frac{\Omega^2}{\gamma}\left(\beta\kappa\sqrt{G(q)/N_1}-1\right) 
\right]^{-1}, &  \mbox{for $\gamma \gg \Omega$,} \\

\left[2\Omega\left(\beta\kappa\sqrt{G(q)/N_1}-1\right)^{1/2} 
\right]^{-1}, & \mbox{for $\Omega \gg \gamma$.}  
\end{array}\right.  
\label{tau2}\end{equation}

Neglecting the contribution of the exchange field to $q$ in the SG phase 
($q\propto (J-T)/J$) one can see the crossover of the temperature dependence 
of the MP formation time at the SG transition. At $\gamma \gg \Omega$ the  
crossover is from $\tau_f\propto T$ in the paramagnetic phase to 
$\tau_f\propto T^{1/2}$ in the SG phase. At $\Omega \gg \gamma$ the crossover  
is from $\tau_f\propto T^{1/2}$ in the paramagnetic phase to $\tau_f\propto
T^{1/4}$ in the SG-phase. The qualitatively different temperature behaviors of 
the MP formation time in the different extreme cases give a possibility to 
infer from the comparison with experiment the relevant corresponding 
dynamical model. The limit case of "strong dissipation" coincides with the 
dynamical model introduced in the papers \cite{{dietl},{20},{21},{22}}. 
 
Taking a typical value of the exciton magnetic polaron formation time of order 
$100ps$ at temperature $T\approx 2 K$ \cite{{dietl},{20}} and the parameters 
$\kappa=41 meV$, $N_1=20$ (the parameters of the solid line in 
Fig.\ref{figmpe}), we can evaluate the order of magnitude of the 
characteristic frequences $\Omega$ and $\gamma$. For the "weak dissipation" 
case ($\Omega \gg \gamma$) we find $\Omega \sim 10^9 Hz$. For the case of 
"strong dissipation" ($\gamma \gg \Omega$) we find $\Omega^2/\gamma\sim 
10^8 Hz$, which implies $\gamma \gg \Omega \gg 10^8 Hz$. The evaluation with 
the parameters $\kappa=65 meV$ and $N_1=50$ corresponding to the dashed-dotted 
line in Fig.\ref{figmpe} would give the same order of the characteristic 
frequencies.

\section{Conclusion}

We elaborated a detailed model description of specific thermodynamic features 
of the magnetic polaron near the spin glass (SG) transition and due to SG 
order. The effect of the exchange interaction between carrier--spin and the  
spin glass is analogous to the effect of a {\it spatially confined exchange 
magnetic field}, which competes with SG order. Consequently, this field leads 
to both a decreased and smeared freezing temperature within the range of the 
magnetic polaron. The onset of spin glass order results in a qualitative 
change of the temperature dependence of the magnetic polaron energy as 
observed experimentally. The obtained solutions of the selfconsistent 
equations for the polaron magnetisation and for the thermodynamically averaged 
hole spin allowed to fit the experimental data of the temperature dependence 
of the exciton magnetic polaron energy in (Cd,Mn)Te. 

An extended model describing different dynamical scenarios of the magnetic 
polaron formation was introduced. These dynamical classes can be distinguished 
by specific thermal behavior of the MP formation time. A comparison with 
experiment on thermal behaviour should hence allow to identify the relevant 
dynamical scenario.

\section{Acknowledgments}
We wish to acknowledge helpful discussions with Dr. Cl. Rigaux, Dr. A. Mauger,  
and collaborators at Universite Paris VI et VII, and with Dr.M. Dahl at 
W\"urzburg university.\\ 
This work was supported by the Deutsche Forschungsgemeinschaft through the 
Sonderforschungsbereich 410.

%\newpage
%\end{multicols}
\newpage 
$\ast$ on leave from A.F. Ioffe Physico--Technical Institute, Russian 
Academy of Sciences, 194021 St. Petersburg, Russia.

%\listoffigures
\psfig{file=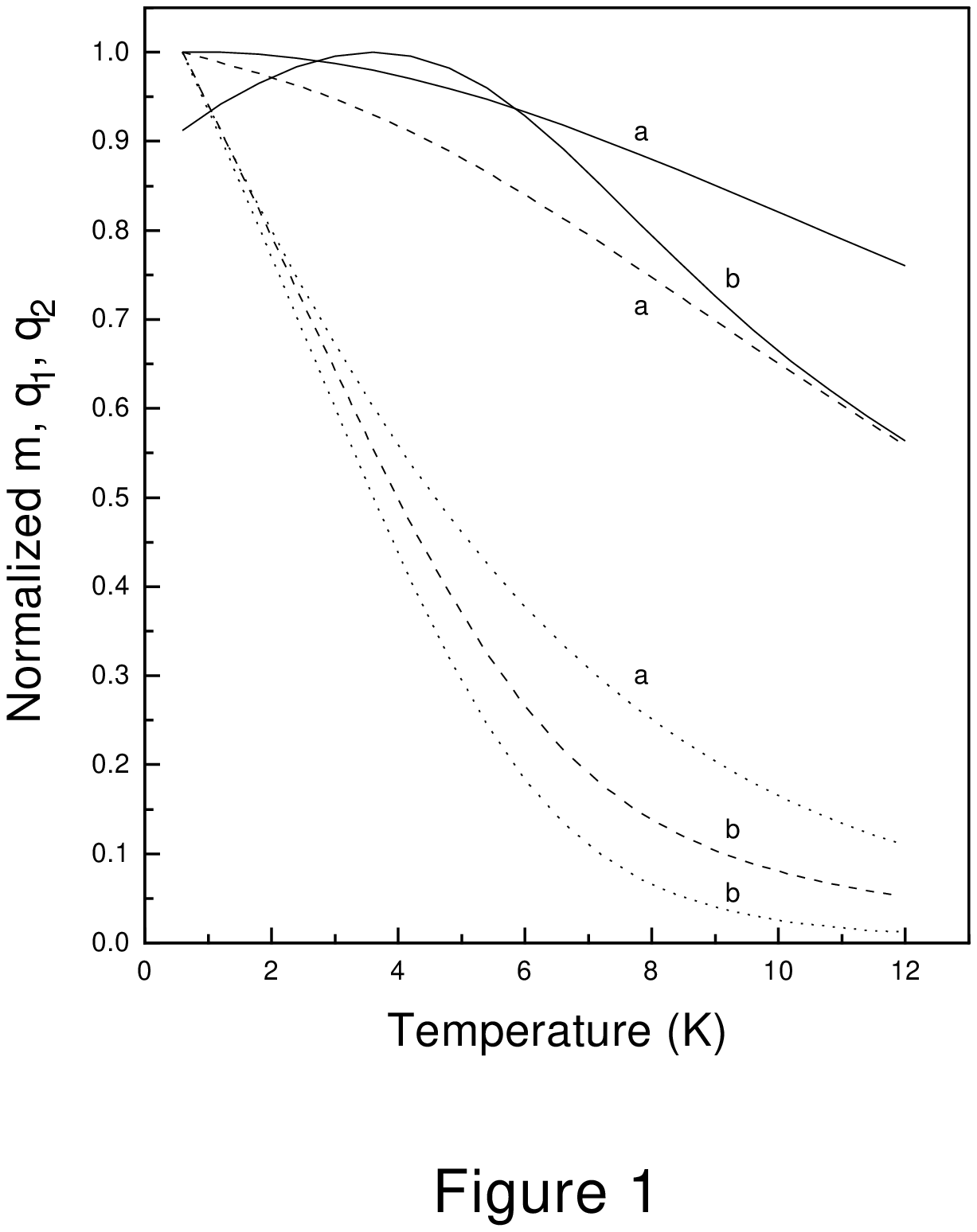,width=16cm,angle=0}
\vspace{-5cm}
\begin{figure}[p]
\caption{Temperature dependences of the magnetization $m$ (solid lines) and  
SG order parameters $q_1$ (dashed lines) and $q_2$ (dotted lines) at two 
different values of exchange constant $\kappa_1=\kappa/N_1$: 
(a) $\kappa_1=3.14 meV$ and (b) $\kappa_1=0.628 meV$. Parameters 
$\kappa=44 meV$ and $\alpha=0.7$ are fixed. The curves are normalized 
to the largest value of the corresponding parameter.  
\label{fig1}}
\end{figure}

\newpage
\psfig{file=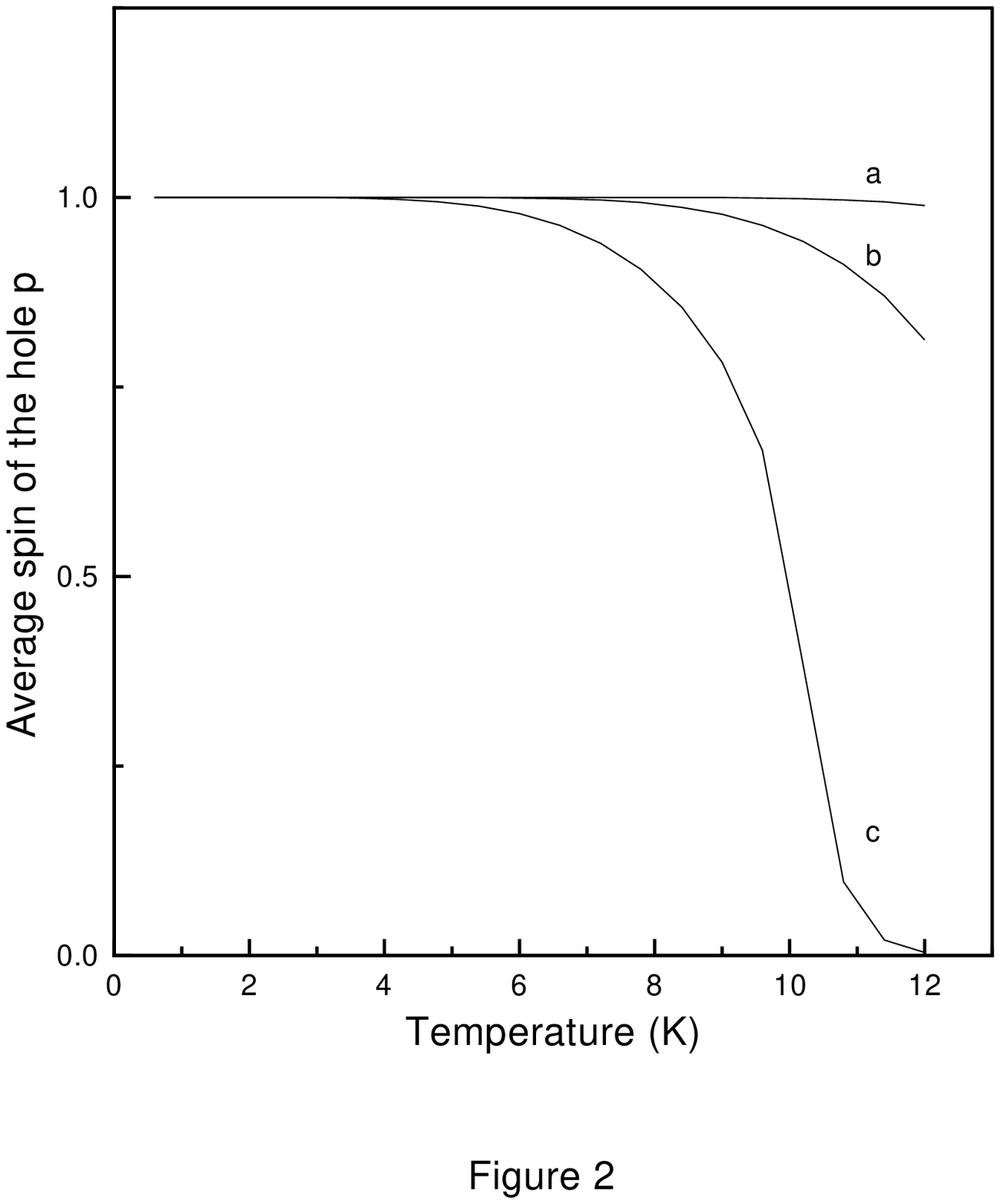,width=16cm,angle=0}
\vspace{-5cm}
\begin{figure}[p] 
\caption{The temperature dependence of the average spin of the hole $p(T)$ 
for different $\kappa$: (a) $\kappa=10 meV$, (b) $\kappa=5 meV$, 
(c) $\kappa=3 meV$. The parameters $J=8 K$, $\alpha=0.7$, $h=0$, and 
$\kappa_1=0.63 meV$ are fixed.
\label{figp-t}}
\end{figure}

\newpage
\psfig{file=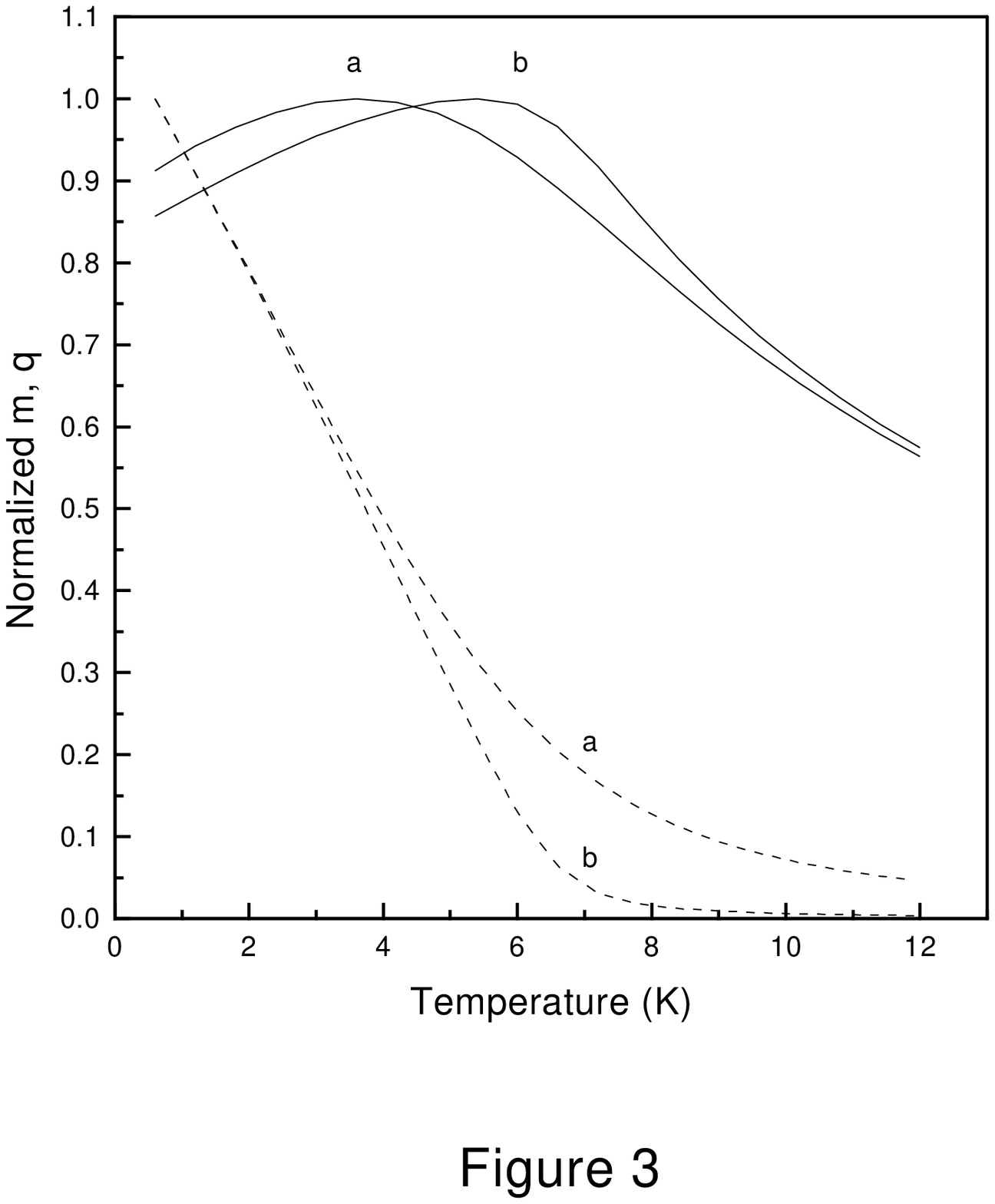,width=16cm,angle=0}
\vspace{-5cm}
\begin{figure}[p] 
\caption{Temperature dependences of the magnetization $m$ (solid lines) and 
the SG order parameter $q=\alpha q_1+(1-\alpha)q_2$ (dashed lines) at two 
different values of $\alpha$: (a) $\alpha=0.7$ and (b) $\alpha=0.2$. Other 
parameters are fixed: $J=8 K$, $\kappa_1=0.628 meV$, $\kappa=44meV$. 
\label{figalpha}}
\end{figure}

\newpage
\psfig{file=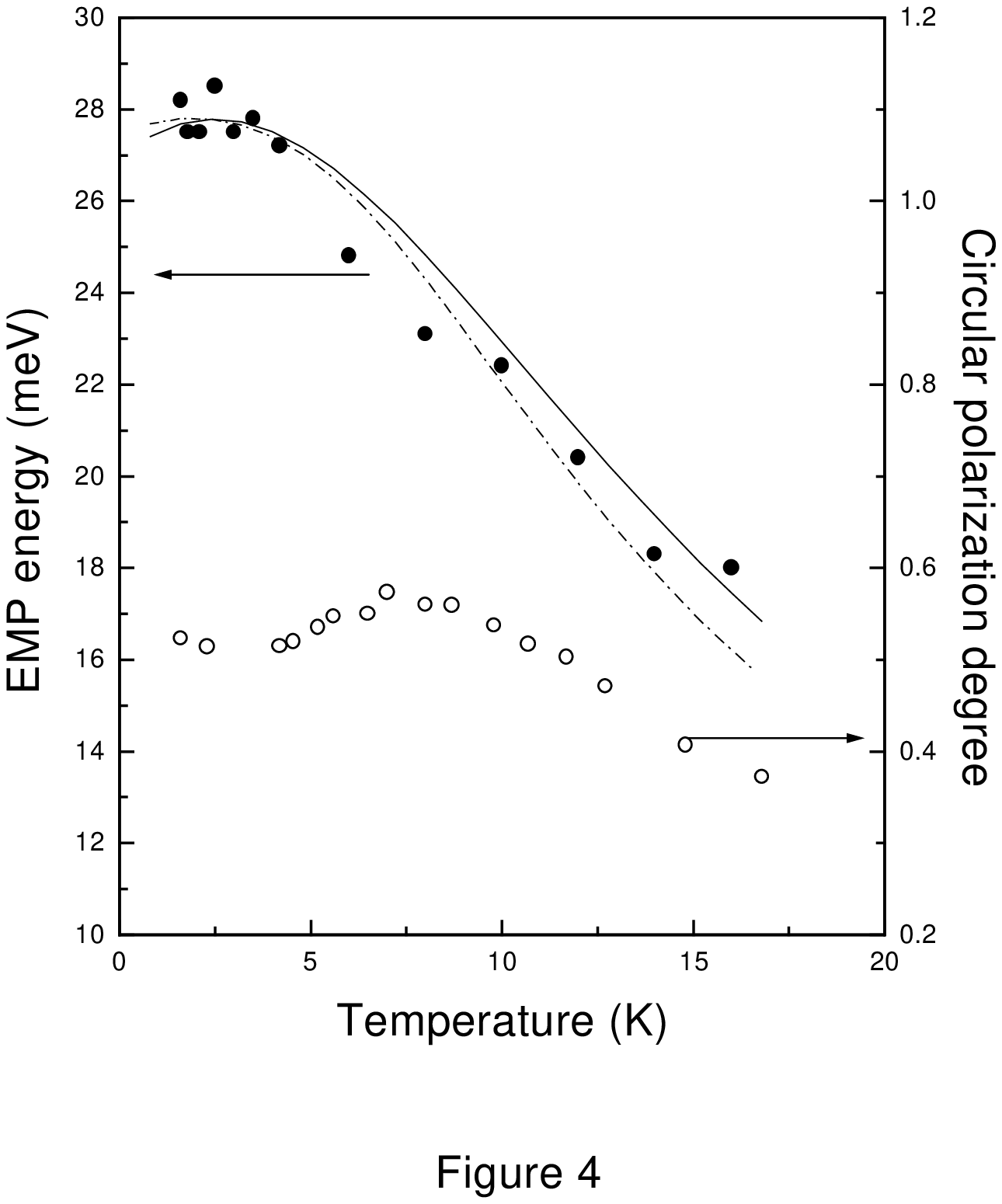,width=16cm,angle=0}
\vspace{-5cm}
\begin{figure}[p]
\caption{The temperature dependence of EMP energy (closed circles) and the 
magnetic-field-induced polarization degree measured at 0.4T (open circles) of 
a MBE-grown epilayer $Cd_{0.67}Mn_{0.33}Te$. Results of theoretical 
calculations: 
Solid line: Fit with SG only, $\kappa=41 meV$, $N=20$, $\alpha=1$.
Dashed-dot line: Fit with SG and small clusters, $\kappa_0=880 meV$, 
$T_{eff}=8 K$, $S_{eff}=0.12$, $\kappa=65 meV$, $N=100$, 
$\alpha=0.5$, $N_1=N_{cl}=50$. 
\label{figmpe}}\end{figure}

\newpage
\psfig{file=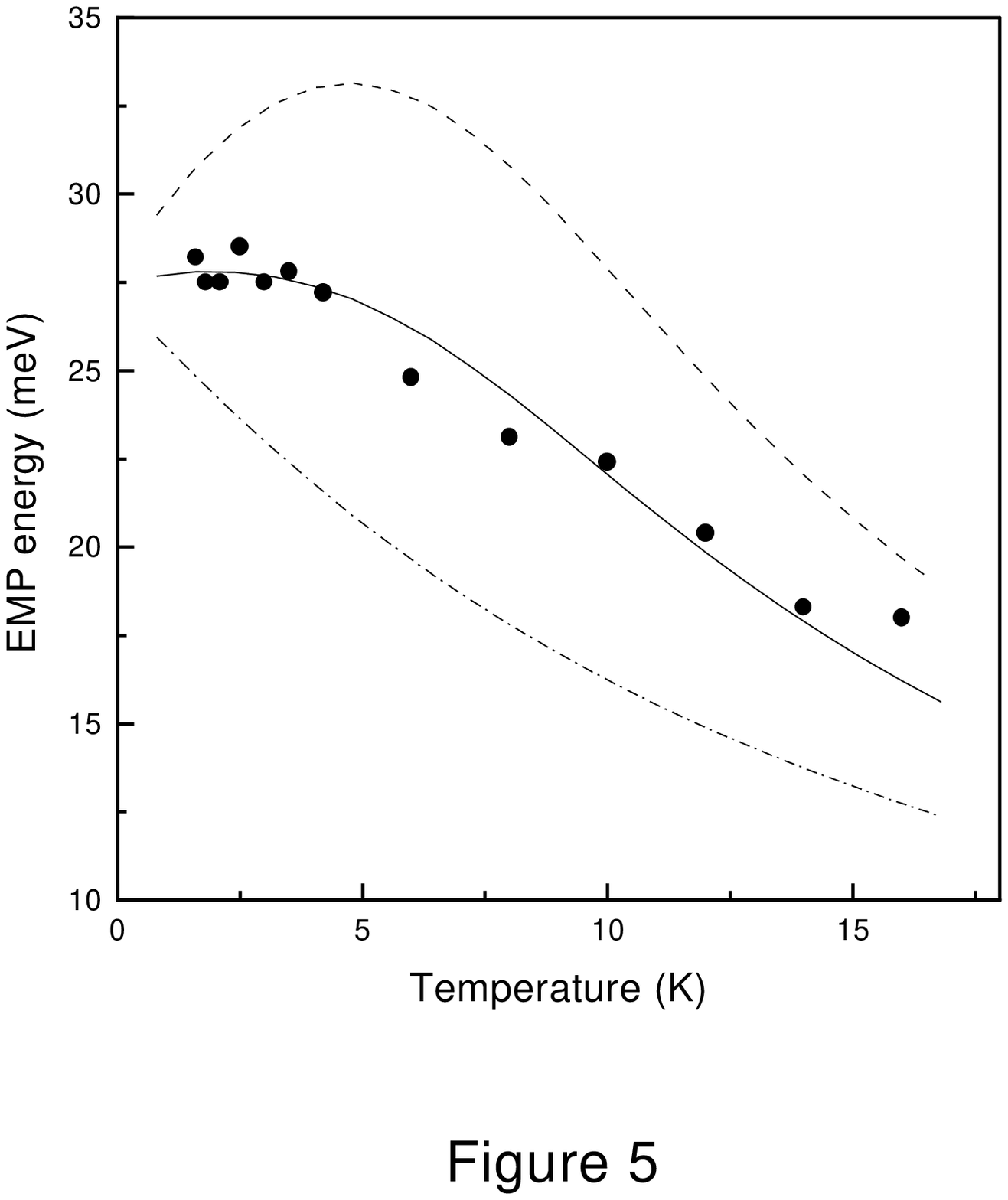,width=16cm,angle=0}
\vspace{-5cm}
\begin{figure}[p]
\caption{The fitting procedure for the model of SG and small spin clusters. 
Solid circles: experimental points as in Fig.\ref{figmpe}. Dashed line: the 
SG contribution to the EMP energy. Dashed-dotted line: the contribution of 
small clusters to the EMP energy. Solid line: the resulting fit curve. The 
parameters of dashed, dashed-dotted and solid lines are the same as by the 
dashed-dotted line in Fig.\ref{figmpe} 
\label{figfit}}
\end{figure}


\begin{thebibliography}{25}
\bibitem{22} The coincidence between the "strong dissipation" scenario and 
the dynamics described in the present paper \cite{20} is obtained if one 
identifies $\Omega^2/\gamma$ in our model with $1/\tau_ S$ in Eq.(1) of 
\cite{20}.
\end{thebibliography}
\end{document}